\documentclass[pra,nopacs,twocolumn,nofootinbib]{revtex4}
\usepackage{graphicx}
\usepackage{bm,amsfonts,amsmath,amssymb}
\usepackage{dcolumn}
\usepackage{natbib}
\usepackage{color,graphicx}

\newcommand{\Eref}[1]{Eq.~(\ref{#1})}
\def\veps{\varepsilon}

\begin{document}

\title{Combination of the Perturbation Theory with Configuration Interaction Method}

\author{M. G. Kozlov$^{1,2}$, I. I. Tupitsyn$^{1,3}$, A. I. Bondarev$^{1,4}$, and D. V. Mironova$^{2}$}

\affiliation{$^1$Petersburg Nuclear Physics Institute of NRC ``Kurchatov Institute'', 188300 Gatchina, Russia}

\affiliation{$^2$St.~Petersburg Electrotechnical University ``LETI'',
Prof.\ Popov Str.\ 5, 197376 St.~Petersburg, Russia}

\affiliation{$^3$Department of Physics, St.\ Petersburg State University, Ulianovskaya 1, Petrodvorets, 198504 St.\ Petersburg, Russia}

\affiliation{$^4$Center for Advanced Studies, Peter the Great St. Petersburg Polytechnic University,
Polytekhnicheskaja 29, 195251 St.\ Petersburg Russia}

\date{\today}

\begin{abstract}
Present atomic theory provides accurate and reliable results for atoms with a small number of valence electrons. However, most current methods of calculations fail when the number of valence electrons exceeds four or five. This means that we can not make reliable predictions for more than a half of the periodic table. Here we suggest a modification of the CI+MBPT (configuration interaction plus many-body perturbation theory) method, which may be applicable to atoms and ions with filling $d$ and $f$ shells. 
\end{abstract}

\maketitle

\section{Introduction}

At present there are several methods of the relativistic correlation calculations of atoms, such as multiconfiguration Dirac-Fock \cite{Des75,JGBF13,FGBJ16},  configuration interaction (CI) \cite{Gu08,FFFG02,JMCB16,Tup03,Tup05}, many-body perturbation theory (MBPT) \cite{DFSS87,BGJS87,Sap98}, CI+MBPT \cite{DFK96,KaBe19,KPST15}, coupled cluster \cite{Eliav2010,Saue-DIRAC20,BJLS89,SaJo08,OZSE20}, and others. Calculations are usually done in the no-virtual-pair approximation using Dirac-Coulomb, or Dirac-Coulomb-Breit approximations \cite{Johnson07}. QED corrections may be included using radiative potential method developed by \citet{FlaGin05}, \cite{GiBe16a}, and QEDMOD potential \cite{STY13,TKSSD16}.  

The coupled cluster method is one of the most popular and effective methods for calculation of atoms with a small number of open shell electrons (or holes). Calculations of the spectra of atoms and ions with many valence electrons (e.~g.\ transition metals, lanthanides, and actinides) are very difficult and usually not very accurate. The reason for that is a combination of strong correlations and a very large configuration space. To account for strong correlations one needs non-perturbative methods, such as CI. On the other hand, a large configuration space makes such calculations very expensive. As a compromise one can try to combine CI with perturbation theory (PT). We will first assume that all closed atomic shells are considered frozen. Then we are treating only valence correlations and consider a combination of the valence CI with valence perturbation theory (VPT). Later we will see that this approach can also be used to treat core-valence correlations.

Recently there were several attempts \cite{DBHF17,GCKB18,DFK19,ImaKoz18} to develop an effective and fast CI+VPT method to speed up calculations for such systems, where straightforward CI calculations are impossible. Application of these methods for systems with a large number of valence electrons was demonstrated in Refs.\ \cite{CSPK20,LiDz20}. A general idea of all these calculation schemes is to make CI in a smaller subspace $P$ and calculate corrections from a complementary subspace $Q$ using VPT. In Refs.\ \cite{DBHF17,GCKB18,DFK19} it is suggested to neglect non-diagonal blocks of the CI matrix in the subspace $Q$, which is equivalent to using VPT. All these methods require summation over all determinants of the complementary subspace $Q$. Though calculating this sum is much easier than calculating and diagonalizing the whole CI matrix, it is still too expensive for the number of valence electrons approaching, or exceeding ten. 

In the paper \cite{DFK19}, the sum over determinants was partly substituted by the sum over configurations that led to a significant increase in calculation speed. Here we want to make another step in this direction. To this end, we will partly substitute VPT with many-body perturbation theory (MBPT). The method we propose here is similar to the old CI+MBPT method \cite{DFK96} but uses different splitting of the problem into the CI and MBPT parts. In particular, we suggest to account for double excitations (D) from the subspace $P$ by means of the MBPT and treat single excitations (S) within VPT, or, if possible, include them directly in CI. We think that this variant is not only more efficient for treating valence correlations, but may also be used for the core-valence correlations.  

\section{Formalism}
\subsection{Valence correlations}

Consider many-electron atom, or ion with $N$ valence electrons, where $N\gg 1$. Let us first assume that other electrons always occupy closed core shells, which is known as a frozen core approximation. Our aim is to solve the $N$-electron Schr\"odinger equation and find the spectrum of this system. 

We start with splitting $N$-electron configuration space in two orthogonal subspaces $P$ and $Q$. The subspace $P$, which we call valence, includes the most important shells. It may be not obvious from the start, which orbitals are `important'. We definitely must include into subspace $P$ all orbitals with occupation numbers of the order of unity in the physical states, we are interested in. Complementary subspace $Q$ includes S, D, and so on excitations from the valence shells to the virtual ones, thus, $Q=Q_S+Q_D+\dots$. We start by solving the matrix equation in the subspace $P$,
\begin{align}\label{eq_CI}
\hat{P}H\hat{P}\Psi_a &= E_a\hat{P}\Psi_a\,,    
\end{align}
where $H$ is the Hamiltonian for valence electrons and $\hat{P}$ is the projector on the subspace $P$. We can find a correction from the complementary subspace $Q$ using the second-order 
perturbation theory:
\begin{align}\label{eq_VPT}
    \delta E_a &= \sum_{n \in Q} \frac{\langle \Psi_a |\hat{P}H\hat{Q}|n\rangle\langle n|\hat{Q}H\hat{P}| \Psi_a \rangle}{E_a-E_n},
\end{align}
where $|n\rangle$ are $N$-electron determinants in the complementary subspace $Q$ and $E_n= \langle n|\hat{Q}H\hat{Q}| n\rangle$.

The wavefunction $\Psi_a$ is a linear combination of the determinants:
\begin{align}\label{eq_Psi}
\Psi_a &= \sum_{m \in P} C^a_m |m\rangle
= \sum_{p,m_p} C^a_{p,m_p} |m_p\rangle
\,.    
\end{align}
Here and below indexes $p$ and $q$ run over configurations in the subspaces $P$ and $Q$ respectively and indexes $m_p$ and $n_q$ numerate determinants within one configuration. Now \Eref{eq_VPT} takes the form: 
\begin{multline}\label{eq_VPTa}
    \delta E_a = 
    \sum_{p,m_p} 
    \sum_{p',m_{p'}} 
    C^a_{p,m_p} C^a_{p',m_{p'}} 
    \\ \times
    \sum_{q,n_q} 
    \frac{\langle m_p |H|n_q\rangle\langle n_q|H| m_{p'} \rangle}{E_a-E_{n_q}},
\end{multline}
where the sum over the subspace $Q$ is also split in two.

For an atom with $N \sim 10$, the dimension of space $Q$ is very large, which makes evaluation of expression \eqref{eq_VPTa} very lengthy. Therefore, our aim is to substitute double sum over $q$ and $n_q$ by a single sum over $q$. To this end we do the following approximation: we substitute the energy $E_{n_q}$ in the denominator by the configuration average:
\begin{align}\label{eq_conf-av}
    \Bar{E}_q &= \frac{1}{N_q}\sum_{n_q=1}^{N_q} E_{n_q}\,,
\end{align}
where $N_q$ is the number of determinants in configuration $q$.
Using this approximation we rewrite \eqref{eq_VPTa} in a form:
\begin{multline}\label{eq_VPTb}
    \delta E_a = 
    \sum_{p,m_p} \sum_{p',m_{p'}}  
    C^a_{p,m_p} C^a_{p',m_{p'}}
    \\
    \times
    \sum_{q} \frac{\langle m_p |H
    \left({\sum_{n_q} |n_q\rangle\langle n_q|}\right) 
    H| m_{p'} \rangle}{E_a-\Bar{E}_q}
    \,.
\end{multline}
Below we will show that in some very important cases one can get rid of the internal sum over $n_q$. 

\begin{figure}[htb]
     \includegraphics[width=0.95\columnwidth]{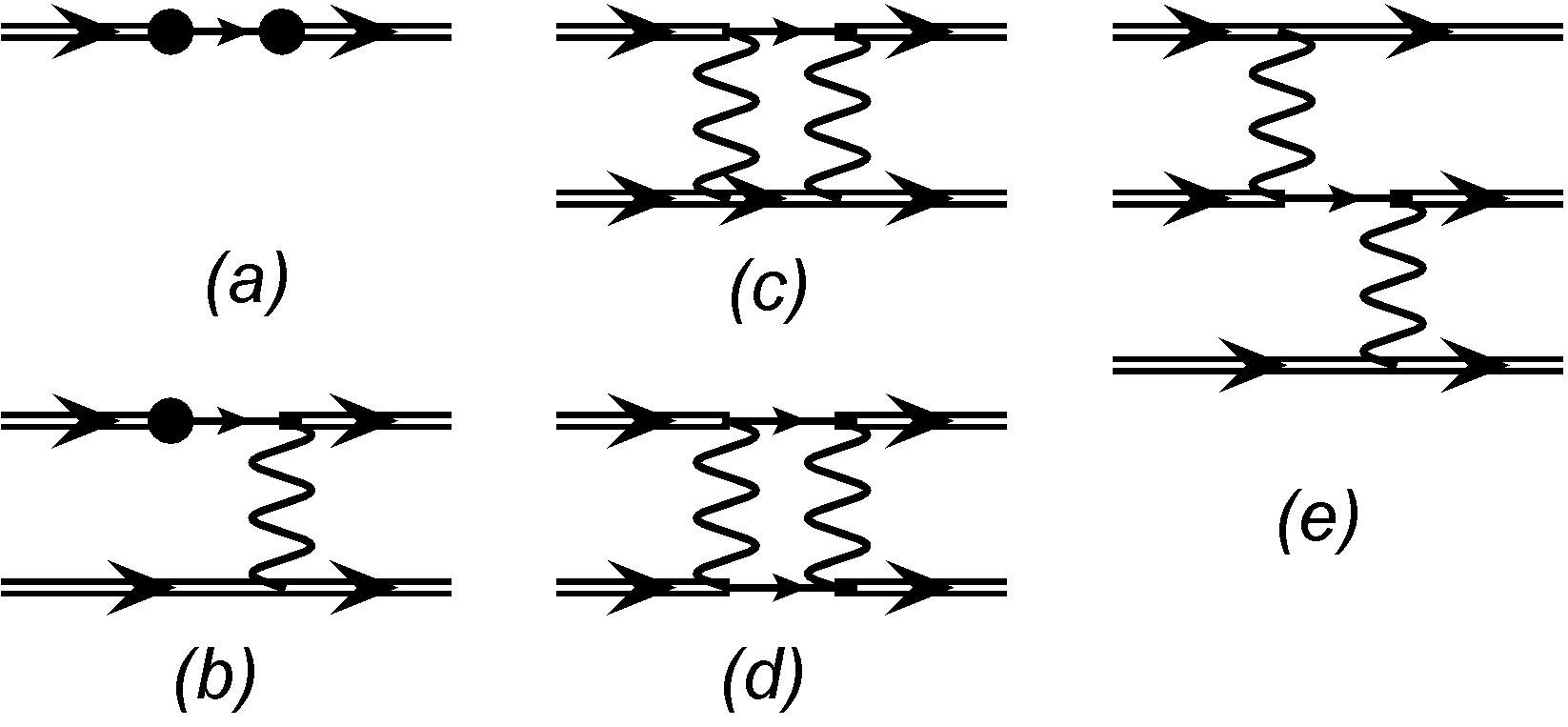}
     \caption{Set of connected second-order diagrams. Black dots correspond to the core potential and wavy lines to the Coulomb interaction. Double and single lines denote electrons in valence and virtual orbitals respectively. Non-symmetric diagrams $(b)$ and $(e)$ have mirror twins. 
     \label{fig:full_set}}
\end{figure}

Hamiltonian $H$ includes one-particle and two-particle parts. The former consists of the kinetic term and the core potential, while the latter corresponds to the Coulomb (or Coulomb-Breit) interaction between valence electrons. Thus, in the sum over $q$ remain only configurations, which differ by no more than two electrons from configurations $p$ and $p'$. This means that within this approximation the subspace $Q$ is actually truncated to $Q_S+Q_D$. All non-zero contributions correspond to the diagrams, shown in Fig.\ \ref{fig:full_set}.

According to our definition of the spaces $P$ and $Q$, the latter must include at least one electron in the virtual shell. Diagrams $(a)$, $(b)$, and $(e)$ include only one intermediate line, so they describe single excitations from the subspace $P$. Diagrams $(c)$ and $(d)$ include two intermediate lines, but only diagram $(d)$ describes double (D) excitations, as both intermediate lines correspond to the virtual shells. 

\begin{figure}[tb]
     \includegraphics[width=0.95\columnwidth]{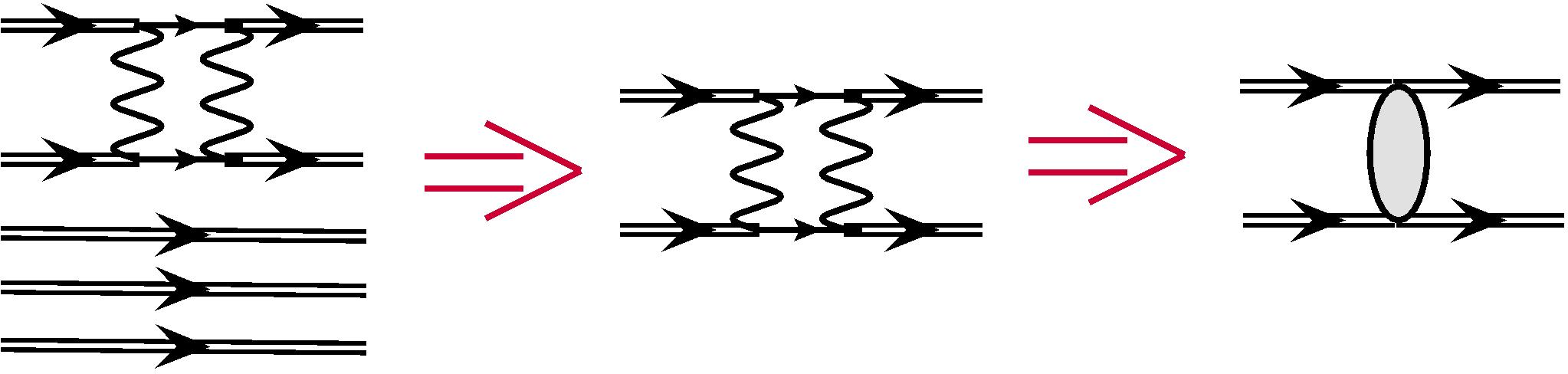}
     \caption{Many-electron second-order expression in \Eref{eq_VPTd} (left) is reduced to the two-particle expression (middle), which, in turn, is reduced to the effective two-particle interaction (right).
     \label{fig:many-el-me}}
\end{figure}

\begin{figure}[htb]
     \includegraphics[width=0.95\columnwidth]{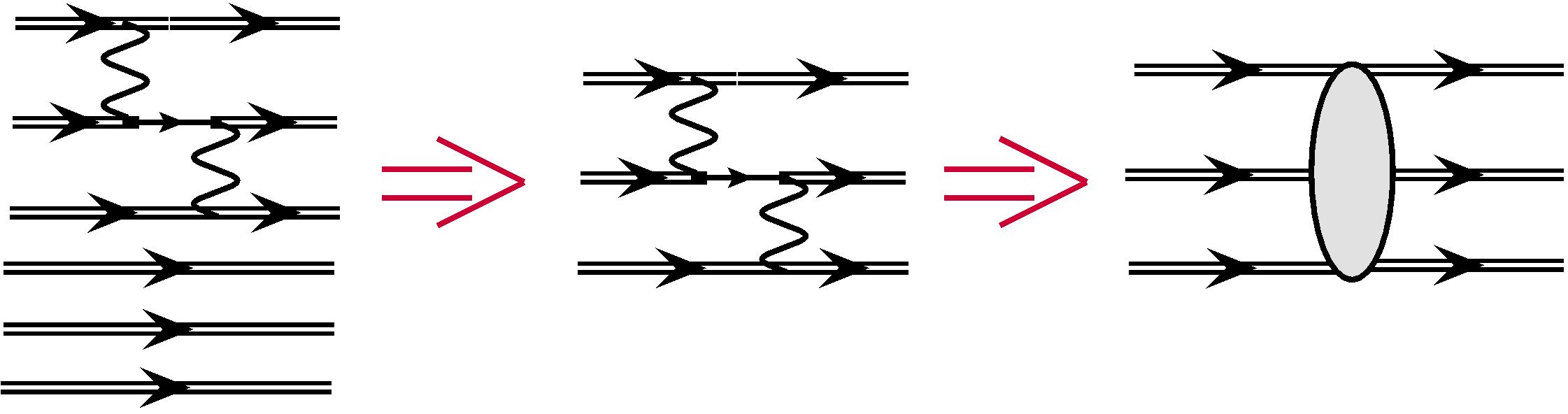}
     \caption{The case when many-electron second-order expression in \Eref{eq_VPTb} (left) is reduced to the three-particle expression (middle), which, in turn, is reduced to the effective three-particle interaction (right). The initial configuration on the left differs from the intermediate configuration by the upper two electrons. The final configuration differs from the intermediate one by the second and third electron from the top. 
     \label{fig:3e-term}}
\end{figure}

Figure \ref{fig:full_set} shows that all many-electron matrix elements in \Eref{eq_VPTb} are reduced to the effective one-electron, two-electron, and three-electron contributions. Effective one-electron contributions are described by diagram $(a)$; diagrams $(b)$, $(c)$, and $(d)$ correspond to the two-electron contributions; finally, diagram $(e)$ describes effective three-electron contributions, see Figs.\ \ref{fig:many-el-me} and \ref{fig:3e-term}. 

For combinatorial reasons the number of configurations with two excited electrons is much bigger, than the number of those with only one such electron. Therefore the vast majority of terms in \Eref{eq_VPTb} correspond to the two-electron excitations from configurations $p$ and $p'$. For these terms in the Hamiltonian $H$ only the two-electron interaction $V$ can contribute, so we can neglect the one-electron part and make substitution $H\to V$. As we saw above, all such terms are described by the single diagram $(d)$ from Fig.\ \ref{fig:full_set}.

Let us consider the sum over doubly excited configurations. It can be written as: 
\begin{multline}\label{eq_VPTd}
    \delta E_a^D = 
    \sum_{p,m_p} \sum_{p',m_{p'}}  
    C^a_{p,m_p} C^a_{p',m_{p'}}
    \\
    \times
    \sum_{q \in Q_D}\frac{\langle m_p |V
    \left({\sum_{n_q} |n_q\rangle\langle n_q|}\right) 
    V| m_{p'} \rangle}{E_a-\Bar{E}_q}
    \,,
\end{multline}
Non-zero contributions come from determinants $|n_q\rangle$, which differ from both determinants $|m_p\rangle$ and $|m_{p'}\rangle$ by two electrons. It is clear that it must be the same two electrons, see Fig.\ \ref{fig:many-el-me}. In this case the second-order expression from \eqref{eq_VPTd} is reduced to the effective two-particle interaction \cite{DFK96}: 
\begin{align}\label{eq_Veff}
    \delta E_a^D &= 
    \sum_{p,m_p} \sum_{p',m_{p'}}  
    C^a_{p,m_p} C^a_{p',m_{p'}}
    \langle m_p |V_\mathrm{eff}| m_{p'} \rangle
    \,.
\end{align}
This effective interaction can be expressed in terms of the effective radial integrals, which are similar to the Coulomb radial integrals. The latter appear when we expand Coulomb interaction in spherical multipoles, 
\begin{align}
V=\sum_{k,\varkappa} V^k_\varkappa.
\end{align}
The matrix element of each multipole component $V^k_\varkappa$ has the form: 
\begin{widetext}
\begin{multline}\label{eq_coulomb_me}
        \langle c,d| V_\varkappa^k |a,b \rangle =
        (-1)^{m_c+m_b+1} \delta_p
        \sqrt{(2j_a+1)(2j_b+1)(2j_c+1)(2j_d+1)}
\\ 
        \left(\!\begin{array}{ccc} j_c & j_a & k\\
        -m_c & m_a & \varkappa \\ \end{array} \!\right)
        \left(\!\begin{array}{ccc} j_b & j_d & k\\
        -m_b & m_d & \varkappa \\ \end{array} \!\right)
        \left(\!\begin{array}{ccc} j_c & j_a & k\\
         \frac12 &-\frac12 & 0 \\ \end{array} \!\right)
        \left(\!\begin{array}{ccc} j_b & j_d & k\\
         \frac12 &-\frac12 & 0 \\ \end{array} \!\right)
        R^k_{a,b,c,d}\,,
\end{multline}
where round brackets denote 3j-symbols, $R^k_{a,b,c,d}$ denotes the Coulomb radial integral and $\delta_p$ ensures parity selection rule: $\delta_p = \xi(l_a+l_c+k) \xi(l_b+l_d+k)$ and $\xi(n) = 1,0$ for $n=\mbox{even, odd}$. A similar multipole expansion holds for the effective interaction $V_\mathrm{eff}$, the effective radial integral being \cite{DFK96}:
\begin{multline}\label{eq_box_diag}
        R^{k,\mathrm{eff}}_{a,b,c,d}
        =\sum_{k_1,k_2}\sum_{m,n} (-1)^\chi
        (2j_m+1)(2j_n+1)(2k+1)
        \left\{\!\begin{array}{ccc} j_c & j_a & k\\
        k_1 & k_2 & j_m \\ \end{array}\! \right\}
        \left\{\!\begin{array}{ccc} j_b & j_d & k\\
        k_2 & k_1 & j_n \\ \end{array} \!\right\}
        \left(\!\begin{array}{ccc} j_m & j_a & k_1\\
         \frac12 &-\frac12 & 0 \\ \end{array} \!\right)
        \left(\!\begin{array}{ccc} j_b & j_n & k_1\\
         \frac12 &-\frac12 & 0 \\ \end{array} \!\right)
\\ 
        \left(\!\begin{array}{ccc} j_c & j_m & k_2\\
         \frac12 &-\frac12 & 0 \\ \end{array} \!\right)
        \left(\!\begin{array}{ccc} j_n & j_d & k_2\\
         \frac12 &-\frac12 & 0 \\ \end{array} \!\right)
        \left(\!\begin{array}{ccc} j_c & j_a & k\\
         \frac12 &-\frac12 & 0 \\ \end{array} \!\right)^{-1}
        \!\!\left(\begin{array}{ccc} j_b & j_d & k\\
         \frac12 &-\frac12 & 0 \\ \end{array} \right)^{-1}
        \frac{R^{k_1}_{a,b,m,n}R^{k_2}_{c,d,m,n}}{\Delta_{E}}\,,
\end{multline}
where curly brackets denote 6j-coefficients, the phase $\chi={j_a+j_b+j_c+j_d+j_m+j_n+k_1+k_2+k+1}$, and $\Delta_E$ is energy denominator, which we will discuss later. For the effective interaction there is no link between parity and multipolarity $k$, so for $V_\mathrm{eff}$ we do not have factor $\delta_p$ as in \Eref{eq_coulomb_me}. The sum in \eqref{eq_box_diag} runs over multipolarities $k_1$ and $k_2$, which satisfies the triangle rule $|k_1-k_2|\le k \le k_1+k_2$. 
\end{widetext}

All single excitations are described by the remaining diagrams from Fig.\ \ref{fig:full_set}. The diagram $(a)$ has a form of the effective one-electron radial integral, while diagrams $(b)$ and $(c)$ are reduced to the two-electron effective radial integrals. In principle, these effective radial integrals can be calculated and stored. However, the diagram $(e)$ corresponds to the effective three-particle interaction. It is difficult to include such interactions into CI matrix for several reasons:
\begin{itemize}
\item When $N> 3$ the number of such effective three-particle integrals is huge.
\item It is difficult to store them and find them.
\item The number of the non-zero matrix elements in the matrix drastically increases. The matrix becomes less sparse and its diagonalization is much more difficult and time-consuming. 
\end{itemize}

Because of all that it is inefficient to use the MBPT approach for three-particle diagrams and it is much easier to treat them within the determinant-based PT. However, it is difficult then to separate them from other contributions, which correspond to single excitations. Thus, it is better not to use MBPT for single excitations at all. We suggest to use instead any form of the determinant-based VPT described in Refs.\ \cite{DBHF17,GCKB18,ImaKoz18,DFK19}. This means that we do VPT in the subspace $Q_S$. Note that the dimension of this subspace is incomparably smaller than the dimension of the $Q_D$ subspace. In some cases it may be so small, that we can include $Q_S$ in the subspace $P$, where we do CI.

\subsection{Core-valence correlations}

\begin{figure}[htb]
     \includegraphics[width=0.95\columnwidth]{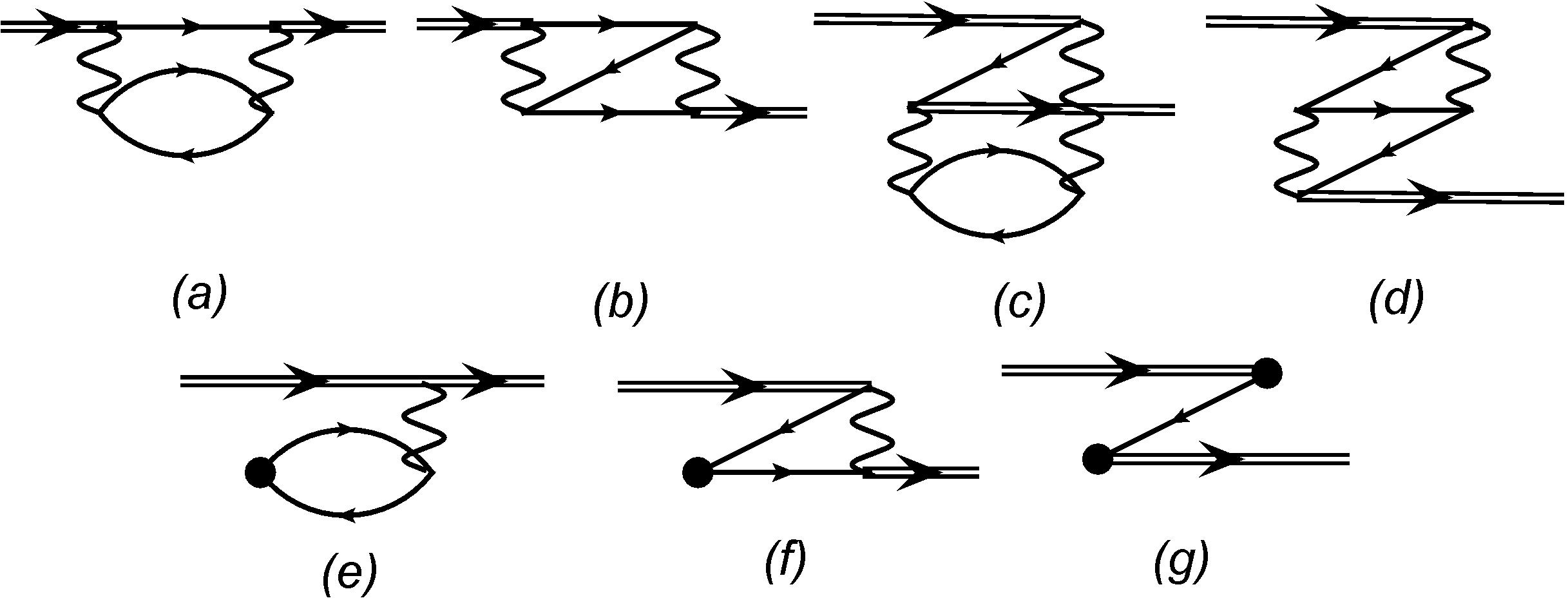}
     \caption{Set of one-electron second-order diagrams accounting for the excitations from the core. Diagrams $(e)$ and $(f)$ have mirror twins. Diagrams $(c)$ and $(d)$ describe double excitations from the core. 
     \label{fig:sigma}}
\end{figure}

\begin{figure}[htb]
     \includegraphics[width=0.95\columnwidth]{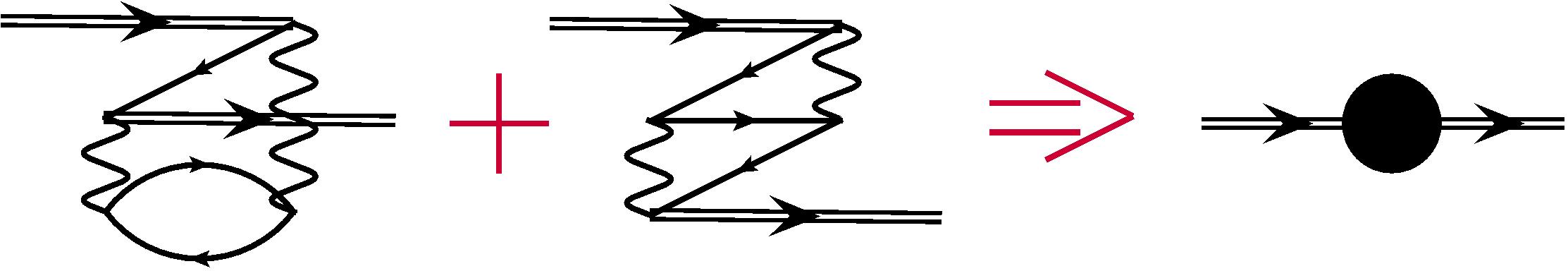}
     \caption{Diagrams, which correspond to the double excitations from closed shells. These diagrams are described by the effective one-electron radial integrals, designated by a black circle. \label{fig:sigma_eff}}
\end{figure}

It is easy to use the scheme described above for the core-valence correlations as well. Now $P$ subspace corresponds to the frozen-core approximation and the subspaces $Q_S$ and $Q_D$ include single and double excitations from the core respectively. This means that these subspaces include many-electron states with one and two holes in the core. As before, the second-order MBPT corrections are described by one-electron, two-electron, and three-electron diagrams. All one-electron diagrams are given in Fig.\ \ref{fig:sigma}. Excitations from the core correspond to the hole lines with arrows looking to the left. It is easy to see that only diagrams $(c)$ and $(d)$ describe double excitations. Therefore, we need to calculate them and store as one-electron effective radial integrals, see Fig.\ \ref{fig:sigma_eff} (note that there are no one-electron contributions for the valence excitations). Expressions for these diagrams were given in Ref.\ \cite{DFK96}.

\begin{figure}[htb]
     \includegraphics[width=0.95\columnwidth]{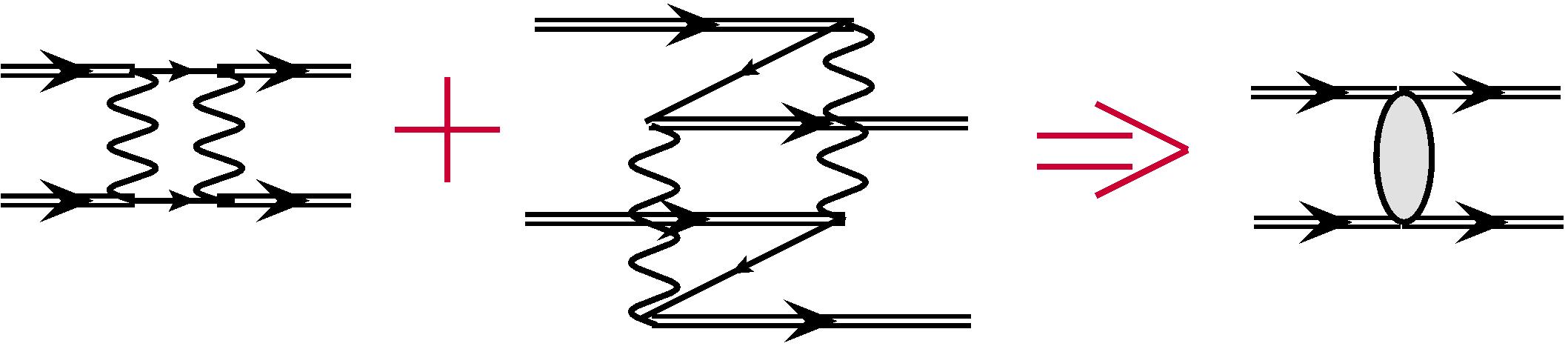}
     \caption{Diagrams contributing to the effective two-electron radial integrals. First diagram accounts for the double excitations to the virtual shells and second diagram accounts for the double excitations from closed shells. \label{fig:R_eff}}
\end{figure}

There is only one two-electron diagram, which corresponds to the double excitations from the core. This diagram must be calculated and added to the similar diagram for valence excitations, which was discussed in the previous section, see Fig.\ \ref{fig:R_eff}. Finally, in analogy with the valence correlations, the three-particle diagrams correspond to the single excitations from the core.

We conclude that in order to account for both valence and core-valence correlations we need to calculate one-electron and two-electron effective radial integrals, which corresponds to the diagrams from Figs.\ \ref{fig:sigma_eff} and \ref{fig:R_eff}. At the same time, we need to include all single excitations from the core shells and all single excitations to the virtual shells either in the subspace $P$ or in the subspace $Q_S$. After that, we make CI calculation with effective radial integrals possibly followed by the VPT calculation in the $Q_S$ subspace. 

\subsection{Sketch of the possible calculation scheme} 

Let us describe a most general computational scheme.
\begin{itemize}
    \item Basis set orbitals are divided into four groups: inner core, outer core, valence, and virtual orbitals. The inner core is kept frozen on all stages of calculation. 
    \item Effective radial integrals are calculated for the valence orbitals, which account for the double excitations from the outer core and the double excitations from the valence orbitals to the virtual ones.
    \item Full CI calculation is done for the valence electrons. The effective radial integrals are added to the conventional radial integrals when the Hamiltonian matrix is formed. 
    \item Determinant-based PT is used in the complementary subspace $Q_S$, which includes single excitations from the outer core and single excitations to the virtual states.
\end{itemize}

Depending on the number of the valence electrons and the size of the core this scheme can be simplified. If there are only two valence electrons, one can include all virtual basis states into valence space. Single excitation from the core can be also added to the valence space. Double excitations from the core are accounted for through the effective radial integrals, while single excitations are included explicitly in the CI matrix. Formally this means that we substitute $P$, $Q$ decomposition by the $P'$, $Q_D$ decomposition:
\begin{align}
\label{PQ}
    &P+Q=P+Q_S+Q_D=P'+Q_D\,,
    \\
\label{PQD}
    &P'\equiv P+Q_S\,.
\end{align}
In the new valence space $P'$, we solve matrix equation with the energy-dependent effective Hamiltonian \cite{DFK96}:
\begin{align}
\label{Heff}
    &H_\mathrm{eff}(E)=H+V_\mathrm{eff}(E)\,,
    \\
\label{HeffPsi}
    &\hat{P'}H_\mathrm{eff}(E_a)\hat{P'}\Psi_a= E_a \hat{P'}\Psi_a\,,
\end{align}
where $\hat{P'}$ is the projector on the subspace $P'$. When the size of the matrix $H_\mathrm{eff}$ becomes too large, one can neglect the non-diagonal part of the matrix in the $Q_S$ space, as in the emu CI method \cite{GCKB18}.

\section{Energy denominators}\label{sec_denom}

Let us discuss the energy denominator $\Delta_E$ in \Eref{eq_box_diag}. For simplicity we will consider the Rayleigh-Schr\"odinger perturbation theory, where the denominator in \Eref{eq_VPTb} would be $\Bar{E}_p-\Bar{E}_q$. Here $\Bar{E}_p$ and $\Bar{E}_q$ are average energies \eqref{eq_conf-av} for configurations $p$ and $q$. Note that in order to return to the Brillouin-Wigner perturbation theory we will need to add $E_a-\Bar{E}_p$, which can be approximately done using the method suggested in \cite{DFK96}. 

In the conventional MBPT the denominator $\Bar{E}_p-\Bar{E}_q$ is reduced to the difference of the Hartree-Fock energies of the orbitals $\veps_i$ which are different in these two configurations. That would give the following energy denominator in \Eref{eq_box_diag}: 
\begin{align} \label{eq_denom}
  \Delta_E \equiv\Delta_E(ab\to mn) = \veps_a +\veps_b -\veps_m -\veps_n\,,  
\end{align}
where we assume that configuration $q$ differs from $p$ by excitation of two electrons from shells $a$ and $b$ to virtual shells $m$ and $n$ respectively. This expression neglects the interaction of the electrons with each other and depends on the choice of the Hartree-Fock potential. In order to improve this approximation, we will consider general expression for the average energy of the relativistic electronic configuration. 

\subsection{Average energy of the relativistic configuration}

The average energy of the relativistic configuration $\Bar{E}_p$ \cite{Mann73,Grant70}:
\begin{multline}\label{eq_Eav}
\Bar{E}_p = 
\sum_{a\in p} q_a \, I_a + \tfrac{1}{2} \,
\sum_{a\in p} q_a\,(q_a-1) \, U_{aa} 
\\
+ \sum_{a < b;\, a,b\in p} q_a \, q_b \, U_{ab},
\end{multline}
where $q_a$ and $q_b$ are occupation numbers for the shells $a$ and $b$ in configuration $p$ and matrix elements of the potential $U$ are given by: 
\begin{align}\label{eq_Uab}
U_{ab} =
\left \{
\begin{array}{ll}
\displaystyle F^0(a,a)+\sum_{k>0}
2 \, f^k_{a,a} \, F^k(a,a) \,,\quad & a=b\,,
\\[5mm]  \displaystyle
F^0(a,b)+\sum_k g^k_{a,b} \, G^k(a,b) \,,\quad & a \ne b\,.
\end{array} \right .
\end{align}
In these equations $I_a$ is the one-electron radial integral, while $F^k(a,b)$ and $G^k(a,b)$ are standard Coulomb and exchange two-electron radial integrals \cite{Grant70}. The angular factors $f^k_{a,a}$ and $g^k_{a,b}$ are also defined in agreement with Ref.\ \cite{Grant70}:
\begin{align}\label{eq_f&g}
\begin{array}{lll}
f^k_{a,a} &= \displaystyle
- \, \frac{1}{2} \,\frac{2j_a+1}{2j_a} \, 
\left (
\begin{array}{llll}
 j_a & j_a & k \\
 \frac{1}{2} & -\frac{1}{2}  & 0
\end{array} \right )^2\,,%
\\[5mm] \displaystyle
g^k_{a,b} &=  \displaystyle
- 
\left (
\begin{array}{llll}
 j_a & j_b & k \\
 \frac{1}{2} & -\frac{1}{2}  & 0
\end{array} \right )^2\,,%
\end{array}
\end{align}
where $j_a$ and $j_b$ are the one-electron total angular momenta.

Let us use \Eref{eq_Eav} to calculate the energy difference between configurations $p$ and $q$ which differ by the excitation of two electrons from shells $a,b$ to shells $m,n$. In other words we need to calculate how the energy changes when occupation numbers change in the following way: $\delta q_a=\delta q_b=-1$ and $\delta q_m=\delta q_n=1$. To this end we, can use Taylor expansion of \Eref{eq_Eav} near the initial configuration $p$: 
\begin{align}\label{eq_Taylor}
\Bar{E}_q=\Bar{E}_p
+ \sum_a \frac{\partial \Bar{E}_p}{\partial q_a} \, \delta q_a
+ \frac{1}{2}  \sum_{a,b} \frac{\partial^2 \Bar{E}_p}{\partial q_a \partial q_b} \,
\delta q_a \, \delta q_b\,,
\end{align}
where derivatives are given by:
\begin{align}
 \frac{\partial \Bar{E}_p}{\partial q_a}
 &= I_a+(q_a-\tfrac12) U_{aa}
 +\sum_{b\neq a}q_b U_{ab}
 \nonumber
 \\
 &= \label{eq_der1}
I_a-\tfrac12 U_{aa}
 +\sum_{b}q_b U_{ab}\,,
\\
 \frac{\partial^2 \Bar{E}_p}{\partial q_a \partial q_b}
 &= \label{eq_der2}
 U_{ab}\,.
\end{align}
Note that all higher derivatives vanish, so expression \eqref{eq_Taylor} is exact. With its help we get:
\begin{multline}\label{eq_ab-mn}
    \Delta_E(ab\to mn)
    = I_a+I_b-I_m-I_n
    \\
    +\sum_{c\in p} q_c\, (U_{ac}+U_{bc}-U_{mc}-U_{nc})
    \\
    -U_{aa}-U_{bb}-U_{ab}-U_{mn}
    \\
    +U_{am}+U_{bn}+U_{an}+U_{bm}\,.
\end{multline}
This expression can be also used for the special cases $a=b,\,\delta q_a=-2$ and/or $m=n,\,\delta q_m=2$. 

Equation \eqref{eq_ab-mn} includes the sum over the occupied shells of the initial configuration $p$. Let us introduce one-electron energies in respect to this configuration as:
\begin{align}\label{eq_eps}
    \veps_a &= I_a +\sum_{c\in p} q_c\, U_{ac}- (1-\delta_{q_a,0})\, U_{aa}\,.
\end{align}
Then \Eref{eq_ab-mn} is simplified to
\begin{multline}\label{eq_ab-mn2}
    \Delta_E(ab\to mn)
    = \veps_a+\veps_b-\veps_m-\veps_n
    \\
    -U_{ab}-U_{mn}+U_{am}+U_{bn}+U_{an}+U_{bm}\,.
\end{multline}
The first line here reproduces the conventional MBPT denominator \eqref{eq_denom}, while the second line gives corrections caused by the interactions of the electrons with each other. It is important that in this form we do not have explicit sums over all electrons, which significantly simplifies calculations. 

In the relativistic calculations the non-relativistic configurations are typically not used. However, sometimes one may need to find the average energy of the non-relativistic configuration. In the Appendix \ref{Appendix_non-rel} we derive the necessary expressions for this case.

\section{Numerical tests}

We made four test calculations for very different systems. In the first two calculations for He I and B I, there was no core and we tested our method for the valence correlations. Then we applied our method for the highly charged ion Fe XVII, where there is a very strong central field, correlation corrections are rather small, and perturbation theory must be quite accurate. In this system we had core $1s^2$, so we calculated core-valence correlation corrections as well as valence ones. Finally, we made calculations for Sc I, where valence $3d$ electrons have a large overlap with the core shell $3p^6$ and core-valence correlation corrections are as important as valence ones.

\begin{table*}[htb]
    \caption{Ground state binding energy of He I (in a.u.). CI calculations are made for three spaces: $P$, $P+Q_S$, and $P+Q$. $\Delta_{P+Q}$ is the difference from the CI result in the $P+Q$ space. Three variants of PT calculations are made based on the CI calculation in $P+Q_S$ space: (a) determinant-based PT; (b) effective Hamiltonian with Hartree-Fock denominators \eqref{eq_denom}; (c) effective Hamiltonian with corrected denominators \eqref{eq_ab-mn2}. Experimental binding energy is given for comparison in the last column \cite{NIST}.}
    \label{tab:HeI}
    \begin{tabular}{lrrrrrrr}
    \hline\hline\\[-8pt]
    &\multicolumn{1}{c}{$P$}
    &\multicolumn{1}{c}{$P+Q_S$}
    &\multicolumn{1}{c}{$P+Q$}
    &\multicolumn{3}{c}{PT}
    &\multicolumn{1}{c}{NIST}\\
    \cline{5-7}
    &&&&\multicolumn{1}{c}{(a)}&\multicolumn{1}{c}{(b)}&\multicolumn{1}{c}{(c)}
    &\multicolumn{1}{c}{Ref.\ \cite{NIST}}\\
    $E(1s^2)$      &$ 2.8626 $&$ 2.8700 $&$ 2.9010 $&$ 2.9021 $&$ 2.9064 $&$ 2.9031 $&$ 2.9034 $ \\
    $\Delta_{P+Q}$ &$ 0.0384 $&$ 0.0310 $&$ 0.0000 $&$-0.0011 $&$-0.0054 $&$-0.0021 $&$-0.0024 $ \\
    \hline\hline
    \end{tabular}
\end{table*}

\begin{table*}[htb]
    \caption{Ground state binding energy of B I (in a.u.). CI calculations are made for valence spaces $P$ and $\tilde{P}$, which included 3 and 4 lower shells respectively.  
    Experimental binding energy is given for comparison in the last column \cite{NIST}.}
    \label{tab:B_I}
    \begin{tabular}{lrrrrrrr}
    \hline\hline\\[-8pt]
    &\multicolumn{1}{c}{$P$}
    &\multicolumn{2}{c}{$P+Q_S$}
    &\multicolumn{1}{c}{$\tilde{P}$}
    &\multicolumn{2}{c}{$\tilde{P}+\tilde{Q}_S$}
    &\multicolumn{1}{c}{NIST}\\
    \cline{3-4} \cline{6-7}
    &&\multicolumn{1}{c}{$H$}&\multicolumn{1}{c}{$H_\mathrm{eff}$}
    &&\multicolumn{1}{c}{$H$}&\multicolumn{1}{c}{$H_\mathrm{eff}$}
    &\multicolumn{1}{c}{Ref.\ \cite{NIST}}
    \\
    $E({}^2P_{1/2})$         &$ 24.5683 $&$ 24.5976 $&$ 24.6595 $&$ 24.5721 $&$ 24.5999 $&$ 24.6581 $&$ 24.6581 $ \\
    $\Delta_{\mathrm{NIST}}$ &$  0.0898 $&$  0.0605 $&$ -0.0014 $&$  0.0860 $&$  0.0582 $&$  0.0000 $&$  0.0000 $ \\
    \hline\hline
    \end{tabular}
\end{table*}

\begin{table*}[htb]
    \caption{Low-lying energy levels of Fe XVII in respect to the ground state (in cm$^{-1}$). The subspace $Q_S$ includes single excitations to virtual shells $n=5-17$. The subspace $Q_S^\prime$ in addition includes single excitations from the $1s$ shell. Effective Hamiltonians account for the respective double excitations. For each calculation we also give relative accuracy in percent.}
    \label{tab:Fe_XVII}
    \begin{tabular}{lcrrrrrrrrrrr}
    \hline\hline\\[-8pt]
    \multicolumn{1}{c}{Config.}&\multicolumn{1}{c}{Level}
    &\multicolumn{1}{c}{NIST}
    &\multicolumn{4}{c}{CI($P$)}
    &\multicolumn{4}{c}{CI$_\mathrm{emu}(P+Q_S)$}
    &\multicolumn{2}{c}{CI$_\mathrm{emu}(P+Q^\prime_S)$}
    \\
    &&\multicolumn{1}{c}{Ref.\ \cite{NIST}}
    &\multicolumn{2}{c}{$H$}    &\multicolumn{2}{c}{$H_\mathrm{eff}$}
    &\multicolumn{2}{c}{$H$}    &\multicolumn{2}{c}{$H_\mathrm{eff}$}    &\multicolumn{2}{c}{$H^\prime_\mathrm{eff}$}
    \\
   \hline
   \\[-8pt]
   \\[-8pt]
 $2p^6   $ & ${}^1S_0  $&$      0$&$      0$&$       $&$      0$&$       $&$      0$&$       $&$      0$&$       $&$      0$&$       $\\
 $2p^5 3p$ & ${}^3S_1  $&$6093450$&$6076370$&$-0.28\%$&$6083540$&$-0.16\%$&$6088405$&$-0.08\%$&$6095600$&$ 0.04\%$&$6095086$&$ 0.03\%$\\
 $2p^5 3p$ & ${}^3D_2  $&$6121690$&$6105049$&$-0.27\%$&$6111933$&$-0.16\%$&$6117307$&$-0.07\%$&$6124215$&$ 0.04\%$&$6123709$&$ 0.03\%$\\
 $2p^5 3p$ & ${}^3D_3  $&$6134730$&$6118010$&$-0.27\%$&$6125056$&$-0.16\%$&$6130067$&$-0.08\%$&$6137137$&$ 0.04\%$&$6136602$&$ 0.03\%$\\
 $2p^5 3p$ & ${}^1P_1  $&$6143850$&$6127278$&$-0.27\%$&$6134193$&$-0.16\%$&$6139345$&$-0.07\%$&$6146283$&$ 0.04\%$&$6145772$&$ 0.03\%$\\[2pt]
 $2p^5 3s$ & $   2^o   $&$5849490$&$5830778$&$-0.32\%$&$5838679$&$-0.18\%$&$5842900$&$-0.11\%$&$5850823$&$ 0.02\%$&$5850330$&$ 0.01\%$\\
 $2p^5 3s$ & $   1^o   $&$5864770$&$5846269$&$-0.32\%$&$5854109$&$-0.18\%$&$5858397$&$-0.11\%$&$5866260$&$ 0.03\%$&$5865678$&$ 0.02\%$\\
 $2p^5 3s$ & $   1^o   $&$5960870$&$5942198$&$-0.31\%$&$5950103$&$-0.18\%$&$5954316$&$-0.11\%$&$5962244$&$ 0.02\%$&$5961601$&$ 0.01\%$\\
 $2p^5 3d$ & ${}^3P_1^o$&$6471800$&$6455306$&$-0.25\%$&$6462010$&$-0.15\%$&$6463149$&$-0.13\%$&$6469882$&$-0.03\%$&$6468962$&$-0.04\%$\\
 $2p^5 3d$ & ${}^3P_2^o$&$6486400$&$6470075$&$-0.25\%$&$6476738$&$-0.15\%$&$6477839$&$-0.13\%$&$6484531$&$-0.03\%$&$6483612$&$-0.04\%$\\
 $2p^5 3d$ & ${}^3F_4^o$&$6486830$&$6471630$&$-0.23\%$&$6478532$&$-0.13\%$&$6478129$&$-0.13\%$&$6485057$&$-0.03\%$&$6484147$&$-0.04\%$\\
 $2p^5 3d$ & ${}^3F_3^o$&$6493030$&$6477585$&$-0.24\%$&$6484338$&$-0.13\%$&$6484319$&$-0.13\%$&$6491101$&$-0.03\%$&$6490177$&$-0.04\%$\\
 $2p^5 3d$ & ${}^1D_2^o$&$6506700$&$6491383$&$-0.24\%$&$6498026$&$-0.13\%$&$6498360$&$-0.13\%$&$6505032$&$-0.03\%$&$6504101$&$-0.04\%$\\
    \hline\hline
    \end{tabular}
\end{table*}

\begin{table*}[htb]
    \caption{Low-lying energy levels of Sc I (in cm$^{-1}$). For each calculation we also give the differences with NIST \cite{NIST} and the average absolute difference $|\Delta|_\mathrm{av} = \frac{1}{k}\sum_{i=1}^k |\Delta_i|$. For the CI calculations in the $P+Q_S$ space we use the emu CI approach \cite{GCKB18} where we neglect non-diagonal matrix elements in the $Q_S$ subspace. On the diagonal we use averaging over relativistic configurations, see \Eref{eq_Eav}.}
    \label{tab:Sc_I}
    \begin{tabular}{lcrrrrrrr}
    \hline\hline\\[-8pt]
    \multicolumn{1}{c}{Config.}&\multicolumn{1}{c}{Level}
    &\multicolumn{1}{c}{NIST}
    &\multicolumn{2}{c}{CI($P$)}
    &\multicolumn{4}{c}{CI$_\mathrm{emu}(P+Q_S)$}
    \\
    &&\multicolumn{1}{c}{Ref.\ \cite{NIST}}
    &\multicolumn{2}{c}{$H$}&\multicolumn{2}{c}{$H$}
    &\multicolumn{2}{c}{$H_\mathrm{eff}$}
    \\
    &&\multicolumn{1}{c}{$E$}&\multicolumn{1}{c}{$E$}&\multicolumn{1}{c}{$\Delta$}
    &\multicolumn{1}{c}{$E$}&\multicolumn{1}{c}{$\Delta$}
    &\multicolumn{1}{c}{$E$}&\multicolumn{1}{c}{$\Delta$}
   \\[2pt]
 $3d 4s^2$ & ${}^2D_{3/2}$ &$     0$&$     0$&$     0$&$     0$&$     0$&$     0$&$     0$\\
 $       $ & ${}^2D_{5/2}$ &$   168$&$   147$&$   -21$&$   157$&$   -11$&$   155$&$   -13$\\[2pt]
 $3d^2 4s$ & ${}^4F_{3/2}$ &$ 11520$&$ 14945$&$  3425$&$  7361$&$ -4159$&$ 11786$&$   266$\\
 $       $ & ${}^4F_{5/2}$ &$ 11558$&$ 14968$&$  3410$&$  7422$&$ -4136$&$ 11847$&$   290$\\
 $       $ & ${}^4F_{7/2}$ &$ 11610$&$ 15001$&$  3391$&$  7489$&$ -4121$&$ 11914$&$   304$\\
 $       $ & ${}^4F_{9/2}$ &$ 11677$&$ 15047$&$  3370$&$  7541$&$ -4136$&$ 11963$&$   285$\\[2pt]
 $3d^2 4s$ & ${}^2F_{5/2}$ &$ 14926$&$ 17368$&$  2442$&$ 11331$&$ -3595$&$ 15661$&$   735$\\
 $       $ & ${}^2F_{7/2}$ &$ 15042$&$ 17455$&$  2413$&$ 11453$&$ -3589$&$ 15781$&$   739$\\[2pt]
 $3d^2 4s$ & ${}^2D_{5/2}$ &$ 17013$&$ 19972$&$  2960$&$ 14574$&$ -2439$&$ 17475$&$   462$\\
 $       $ & ${}^2D_{3/2}$ &$ 17025$&$ 19980$&$  2955$&$ 14601$&$ -2424$&$ 17500$&$   475$\\[2pt]
 $3d^2 4s$ & ${}^4P_{1/2}$ &$ 17226$&$ 20329$&$  3103$&$ 14606$&$ -2620$&$ 17472$&$   246$\\
 $       $ & ${}^4P_{3/2}$ &$ 17255$&$ 20339$&$  3084$&$ 14679$&$ -2576$&$ 17552$&$   297$\\
 $       $ & ${}^4P_{5/2}$ &$ 17307$&$\quad 20380$&$  3073$&$\quad 14739$&$ -2568$&$\quad 17606$&$   299$\\[2pt]
 $3d4s4p $ & ${}^4F_{3/2}^o$ &$ 15673$&$ 13921$&$ -1751$&$ 16019$&$   346$&$ 15872$&$   200$\\
 $       $ & ${}^4F_{5/2}^o$ &$ 15757$&$ 14002$&$ -1754$&$ 16099$&$   342$&$ 15953$&$   197$\\
 $       $ & ${}^4F_{7/2}^o$ &$ 15882$&$ 14139$&$ -1743$&$ 16211$&$   330$&$ 16064$&$   183$\\
 $       $ & ${}^4F_{9/2}^o$ &$ 16027$&$ 14290$&$ -1737$&$ 16340$&$   314$&$ 16194$&$   168$\\[2pt]
 $3d4s4p $ & ${}^4D_{1/2}^o$ &$ 16010$&$ 14265$&$ -1745$&$ 16318$&$   308$&$ 16448$&$   438$\\
 $       $ & ${}^4D_{3/2}^o$ &$ 16022$&$ 14311$&$ -1711$&$ 16351$&$   329$&$ 16517$&$   495$\\
 $       $ & ${}^4D_{5/2}^o$ &$ 16141$&$ 14375$&$ -1766$&$ 16403$&$   262$&$ 16559$&$   418$\\
 $       $ & ${}^4D_{7/2}^o$ &$ 16211$&$ 14458$&$ -1753$&$ 16503$&$   292$&$ 16621$&$   410$\\[2pt]
 $3d4s4p $ & ${}^2D_{3/2}^o$ &$ 16023$&$ 14172$&$ -1851$&$ 16516$&$   493$&$ 16442$&$   419$\\
 $       $ & ${}^2D_{5/2}^o$ &$ 16097$&$ 14189$&$ -1907$&$ 16525$&$   428$&$ 16449$&$   352$\\[2pt]
 $3d4s4p $ & ${}^4P_{1/2}^o$ &$ 18504$&$ 16854$&$ -1650$&$ 18528$&$    24$&$ 18529$&$    25$\\
 $       $ & ${}^4P_{3/2}^o$ &$ 18516$&$ 16930$&$ -1586$&$ 18538$&$    23$&$ 18543$&$    27$\\
 $       $ & ${}^4P_{5/2}^o$ &$ 18571$&$ 17007$&$ -1565$&$ 18577$&$     6$&$ 18572$&$     1$\\[2pt]
 \multicolumn{2}{c}{$|\Delta|_\mathrm{av}$}&
 &\multicolumn{2}{c}{2247}&\multicolumn{2}{c}{1595}&\multicolumn{2}{c}{310}
\\
    \hline\hline
    \end{tabular}
\end{table*}

\subsection{Ground state of He I}

Helium is the simplest system where correlation effects can be tested. We calculate the ground state energy, where correlation corrections are the largest. We choose the space $P$ to include shells $n=1\dots3$. The space $Q$ includes virtual shells $s,p,d$ with $4\le n\le20$. For this model problem, we can easily do CI in the whole space $P+Q$ thus producing the ``exact'' solution and compare these results with different variants of the perturbation theory discussed above. Results are listed in Table \ref{tab:HeI}.

One can see that the valence CI provides accuracy on the order of 1\%. The accuracy does not improve when we account for the single excitations to the virtual shells. However, when we include double excitations the agreement with the ``exact'' answer is significantly better. The determinant-based PT gives the best result. The results obtained with the effective Hamiltonian are less accurate, but corrections to the denominators reduce the discrepancy. Even the uncorrected variant of the MBPT is closer to the ``exact'' answer by an order of magnitude compared to the valence CI. 

\subsection{Ground state of B I}
B is a five electron system. The full CI calculation here is already very expensive. The determinant-based PT is also rather lengthy, so we made calculations only with the effective Hamiltonian and compared our results with the experiment \cite{NIST}. The effective radial integrals were calculated using the Hartree-Fock denominators. We tested two variants of the valence space: the first one, $P$, included shells $n=1\dots 3$ and the second one, $\tilde{P}$, included also the shell $n=4$. Corresponding $Q$ and $\tilde{Q}$ spaces included $s,p,d,f,g$ shells up to $n=20$. Results of these calculations for the ground state ${}^2P_{1/2}$ are given in Table \ref{tab:B_I}. We see that the accuracy of the CI calculation does not change much when we include an extra shell in the subspace $P$. The accuracy of the CI calculation in the subspaces $P+Q_S$ and $\tilde{P}+\tilde{Q}_S$ is only slightly better than similar calculation in the subspaces $P$ and $\tilde{P}$. Only including double excitations by means of the MBPT improves the agreement with the experiment by more than an order of magnitude.  

\subsection{Spectrum of Fe XVII}

Ten-electron ion Fe XVII plays an important role in astrophysics and plasma physics, see Ref. \cite{Kuhn2020} and references therein. The spectrum of this ion was calculated within several different approaches \cite{Kuhn2020suppl} with relative accuracy of about 0.03\%. Here we repeat these calculations using the new method. We use basis set $[17spdfg]$. Virtual orbitals starting from $4s$ and up are formed from B-splines using the method from Ref.\ \cite{KozTup19}. Valence subspace $P$ includes shells $2s,2p,3s,3p,3d,4s,4p,4d$, and $4f$, while the $1s$ shell is frozen. Single excitations to all higher orbitals are included in the subspace $Q_S$ and the subspace $Q_S^\prime$ in addition includes single excitations from the $1s$ shell. We make two CI calculations in the spaces $P$ and $P+Q_S$ respectively. Then we repeat these calculations using the effective Hamiltonian, which accounts for the excitations to the subspace $Q_D$. Finally, we make CI calculation in the $P+Q_S^\prime$ for the effective Hamiltonian $H_\mathrm{eff}^\prime$ which accounts for the double excitations from $1s$ shell as well as for the double excitations to the virtual shells with $n \ge 5$. Results of all these calculations are given in Table \ref{tab:Fe_XVII}.

One can see that already the CI calculation in the subspace $P$ is quite accurate here, the relative errors being about 0.3\%. This is not surprising for such a strong central field. When we increase the size of the configuration space by adding single excitations to the virtual shells $n=5\dots 17$ the errors substantially decrease but remain of the same order of magnitude. The same happens when we do CI for the effective Hamiltonian in the subspace $P$. Only when we include both single and double excitations to the virtual shells by doing CI for the effective Hamiltonian in the subspace $P+Q_S$ we increase the accuracy by an order of magnitude, the errors being  0.04\% or less. Adding S and D excitations from the $1s$ shell leads to corrections to the transition energies within $0.01\%$. Our final accuracy is similar to the accuracy obtained in Ref.\ \cite{Kuhn2020suppl}, where CI space included all double and some triple excitations to all virtual shells (the basis set there was different, but of the same length). In our present calculation, the size of the space $P+Q_S$ is about 1.4 million determinants, and the size of the space $P+Q_S^\prime$ is close to 2 million determinants, which is significantly less than the CI space of Ref.~\cite{Kuhn2020}.

\subsection{Spectrum of Sc I}

The ground state configuration for Sc I is $\mathrm{[Ar]} 3d^1 4s^2$ and lowest excited states belong to the configurations $3d^2 4s$ and $3d 4s 4p$. The $3d$ shell has a large overlap with the core shells $3s$ and $3p$. Because of that frozen core approximation can not reproduce even the lowest part of the spectrum. Including $3s$ and $3p$ shells into the valence space makes its size extremely large. Therefore, this is a good system to apply our method. 

We use a short basis set $[9spdfgh]$, which is constructed as described in Ref.\ \cite{KozTup19}. In the valence space $P$, the shells $n\le 3$ are closed and the virtual shells $n\ge 8$ and all $h$ orbitals are empty. The space $Q_S$ includes single excitations from the upper core shells $n=3$ and single excitations to the virtual shells. We keep core shells up to $n\le 2$ frozen on all stages. Results of the calculation of the spectrum are presented in Table \ref{tab:Sc_I}, where excitation energies from the ground state in cm $^{-1}$ are shown for approximately 10 lower levels of each parity. The sizes of the valence space $P$ and $P+Q_S$ are about $6\times 10^4$ and $1\times 10^6$ determinants respectively. We list the results of three calculations: the full CI in the valence space $P$ and emu CI \cite{GCKB18} in the space $P+Q_S$ for the bare and the effective Hamiltonians. The effective radial integrals were calculated with the Hartree-Fock denominators. For each of these calculations we also give differences from the experimental values \cite{NIST} and the averaged absolute difference. 

One can see that all the levels in the CI calculation are shifted from their experimental energies:  the levels of the configuration $3d^2 4s$ lie higher by 3 thousand inverse centimeters, while the levels of the configuration $3d4s4p$ lie lower by 2 thousand inverse centimeters. The picture changes drastically when we add single excitations and solve the problem in the space $P+Q_S$. Now the levels of the configuration $3d^2 4s$ lie lower by 3 thousand inverse centimeters, while the levels of the configuration $3d4s4p$ are almost in place. Finally, when we use the effective Hamiltonian, which accounts for the double excitations, the levels get closer to their places with the average deviation about 300 cm$^{-1}$, or 7 times smaller, than for the CI calculation. 

In this test calculation, we used a rather short basis set and were probably rather far from saturation. Therefore we can not reliably estimate the ultimate accuracy of the method for scandium. Looking at the results we see that the size of the PT corrections is very large and there is also large cancellation between contributions of the single and double excitations. Therefore it is unlikely that converged results would be significantly better than what we got here. On the other hand, we see systematic improvement in our final results compared to the pure valence calculation. It is also worth mentioning that if one would try to include all double excitations in CI calculation the size of the configuration space would be much above $1\times 10^8$ even for the basis set as short as this one.  

\section{Conclusions}

We suggest a new version of the CI+MBPT method \cite{DFK96} with the different division of the many-electron space into parts where non-perturbative and perturbative methods are used. This new division may be more practical for the atoms with many valence electrons, where the size of the valence space may be too big for solving the matrix eigenvalue problem. This method can be used in the all-electron calculations for light atoms as well as for the calculations with the frozen core. In the latter case, the single and double excitations from (some of) the core shells can be treated perturbatively. We ran four rather different tests which showed systematic one-order-of-magnitude improvement of the results when we added MBPT corrections to the CI calculations.

\section*{Acknowledgements}

We thank Marianna Safronova, Charles Cheung, and Sergey Porsev for their constant interest in this work and very useful discussions. This work was supported by the Russian Science Foundation (Grant No. 19-12-00157). I.I.T. acknowledges the support from the Resource Center ``Computer Center of SPbU'', St. Petersburg, Russia. 
\\

\appendix
\section{Average energy of the non-relativistic configuration}
\label{Appendix_non-rel}

In the average over non-relativistic configuration (LS-average)~\cite{TZSP18,LinRos75}, the occupation numbers for the relativistic orbitals $q_a$ may be non-integer, while occupation numbers for non-relativistic orbitals $q_A$ are still integer (we use capital letters $A,B,M,N$ to designate non-relativistic orbitals). Below we show that properly defining one-electron integrals $I_A$ and two-electron matrix elements $U_{AB}$ we obtain expressions similar to Eqs.\ (\ref{eq_eps},~\ref{eq_ab-mn2}).

The average energy of the non-relativistic configuration $R$ can be written as: 
\begin{widetext}
\begin{multline}
\bar{E}_R = \sum_{a} \tilde q_a \, I_a+
\frac{1}{2}\sum_{a} \tilde q_a \, (\tilde q_a-w_a) \, F^{0}(a,a)
+\sum_{a<b}  \tilde q_a \, \tilde q_b \,w_{ab} \, F^{0}(a,b)
\\
+ \sum_{a,k>0}  \tilde q_a ( \tilde q_a-w_a) \, f^k_{aa} \, F^k(a,a)
 +\sum_{a<b,k}  \tilde q_a \,  \tilde q_b \, w_{ab} \, g^k_{ab} \, G^k(a,b)\,,
\label{etot1}
\end{multline}
where
\begin{align}
&\tilde q_a = \displaystyle \frac{2j_a+1}{4l_a+2} \, q_{A} \,,
\qquad
w_a = \displaystyle \frac{q_{A}-\tilde q_a+2j_a}{4l_a+1}\,,
\qquad
w_{ab} = \left \{
\begin{array}{cll} 
\frac{4l_a+2}{4l_a+1} \, \frac{q_A-1}{q_A} \,, & \, A=B\,,
\\
1 \,, & \, A \ne B\,.
\end{array} \right .
\end{align}
Using expressions:
\begin{align}
&\tilde q_a (\tilde q_a -w_a) = \frac{2j_a+1}{4l_a+2} \,
\frac{2j_a}{4l_a+1} \,q_{A} (q_A-1) \,,
&w_{ab} \, \tilde q_a \tilde q_b = \frac{2j_a+1}{4l_a+2} \,
\frac{2j_{a^\prime}+1}{4l_a+1} \,q_{A} (q_{A}-1) \,, 
\quad A=B,
\,\, j_a \ne j_{a^\prime}\,,
\end{align}
we rewrite the equation (\ref{etot1}) in the form
\begin{multline}
\bar{E}_R = \sum_{A} q_A \,\sum_{j_a}\frac{2j_a+1}{4l_a+2} I_a+
\frac{1}{2}\sum_{A} q_A \, (q_A-1) \, \sum_{a,a^\prime \in A}
\frac{(2j_a+1)(2j_{a^{\prime}}+1-\delta_{a,a^{\prime}})}{(4l_a+2)(4l_a+1)} \,
F^{0}(a,a^{\prime})+
\\
+\frac{1}{2} \sum_{A \ne B} q_{A} \, q_{B} \sum_{a \in A, b \in B}
\frac{(2j_a+1)(2j_b+1)}{(4l_a+2)(4l_b+2)}\,
\left[F^{0}(a,b)+ \sum_k  g^k_{ab} \, G^k(a,b) \right]
\\
+\frac{1}{2} \sum_{A} \,q_{A} (q_{A}-1) \sum_{a,a^{\prime} \in A} \sum_{k>0} \,
\frac{2j_a+1}{4l_a+2} \, \frac{2j_{a^\prime}+1}{4l_a+1} \,
g^k_{aa^{\prime}} \, G^k(a,a^\prime) \,.
\label{etot2}
\end{multline}
In the last sum, the term $k=0$ is absent since $j_a \ne j_{a'}$ and
$k \ge |j_a-j_{a'}|$.
Now we can introduce non-relativistic analogues of the integrals $I_a$ and matrix elements $U_{ab}$ and rewrite \Eref{etot1} alike \Eref{eq_Eav}:
\begin{align}
\bar{E}_R &= \sum_{A} q_{A} \, I_{A} + \frac{1}{2} \,
\sum_{A} q_{A} \, (q_A-1) \, U_{AA} +
\frac{1}{2} \, \sum_{A \ne B} \, q_{A} \, q_{B} \, U_{AB} \,,
\label{etot3}
\\
I_A &= \sum_{a \in A} \frac{2j_a+1}{4l_a+2} \, I_a\,,
\\
U_{AA} &= \sum_{a,a^\prime \in A} 
\frac{(2j_a+1)(2j_{a^{\prime}}+1)}{(4l_a+2)(4l_a+1)} \, \left[ F^{0}(a,a^{\prime})+
\sum_{k>0}  g^k_{aa^{\prime}} \, G^k(a,a^{\prime}) \right ] -
\sum_{a \in A} \frac{ (2j_a+1)}{(4l_a+2)(4l_a+1)} \, F^{0}(a,a)\,,
\\
U_{AB} &= \sum_{a \in A, b \in B}
\frac{(2j_a+1)(2j_b+1)}{(4l_a+2)(4l_b+2)}\,
\left[F^{0}(a,b)+ \sum_k  g^k_{ab} \, G^k(a,b) \right]
\,.
\end{align}

Using \Eref{etot3} we get the following derivatives by analogy with Eqs. (\ref{eq_der1},~\ref{eq_der2}): 
\begin{align}
&\frac{\partial \bar{E}_R}{\partial q_A} 
= I_A+
\sum_{B} q_{B} \, U_{A B}-\frac{1}{2} \, U_{AA} \,,
&\frac{\partial^2 \bar{E}_R}{\partial q_A \partial q_B} &= \displaystyle U_{AB} \,.
\end{align}
The difference in energy between two configurations is
\begin{align}
\Delta \bar{E} &= \sum_A I_A \, \delta q_A  +
\sum_A \left(q_A - \frac{1}{2} \right) \, U_{AA} \, \delta q_A
+\sum_{B \ne A} q_B \, U_{AB} \, \delta q_B +
\frac{1}{2}  \sum_{A, B} U_{AB} \, \delta q_A \, \delta q_B \,.
\end{align}
This equation allow us to find the energy of a double excitation 
$\delta q_A =-1,\, \delta q_B=-1,\, \delta q_N=1,\,
\delta q_{M}=1$:
\begin{multline}\label{eq_dbl_ex}
\Delta_{\bar{E}}(AB \to NM) =  
I_A + I_B- I_M - I_N 
+\sum_{C} q_C \, (U_{AC}+U_{BC}-U_{MC}-U_{NC})
\\
- U_{AA}-U_{BB}-U_{AB}-U_{NM}
+U_{AN}+U_{BN}+U_{AM}+U_{BM}\,.
\end{multline}
If we introduce an averaged one-electron energy by analogy with \eqref{eq_eps} we can rewrite \eqref{eq_dbl_ex} as
\begin{align}
\label{eq_hf_energy_av}
\varepsilon_A &= I_A + \sum_{B} q_B \, U_{AB} -
(1-\delta_{q_A,0}) \,U_{AA}\,,
\\
\label{eq_dbl_ex2}
\Delta_{\bar{E}}(AB \to NM)   
&=\varepsilon_A + \varepsilon_B- \varepsilon_M -\varepsilon_N
-U_{AB}-U_{NM}+U_{AN}+U_{BN}+U_{AM}+U_{BM} \,.
\end{align}

We obtained corrections to the standard MBPT energy denominator \eqref{eq_denom} using two approximations. Averaging over relativistic configurations gives expression \eqref{eq_ab-mn2} and averaging over non-relativistic configurations leads to expression \eqref{eq_dbl_ex2}. These expressions differ only by the definitions of the one-electron energies and two-electron matrix elements.  
\end{widetext}%


\end{document}